\documentclass{PoS}
\title{Width of the Confining String}
\ShortTitle{Width of the Confining String}
\author{\speaker{F.V.~Gubarev}\thanks{Work was supported by RFBR-07-02-00237-a,
				      RFBR-08-02-00661-a grants, computations
				      had been partially performed at Joint
				      Supercomputer Center, RAS.}\\
	Institute of Theoretical and  Experimental Physics, \\
        B.~Cheremushkinskaya 25, Moscow, 117218, Russia\\
        E-mail: \email{gubarev@itep.ru}}
\usepackage{latexsym}
\usepackage{psfig}
\usepackage{multirow}
\newcommand{\beq}{\begin{equation}}
\newcommand{\eeq}{\end{equation}}
\newcommand{\beqn}{\begin{eqnarray}}
\newcommand{\eeqn}{\end{eqnarray}}
\newcommand{\bea}[1]{\beq\begin{array}{#1}}
\newcommand{\eea}{\end{array}\eeq}

\newcommand{\tr}{\mathop{\rm Tr}}

\newcommand{\ket}[1]{|\,#1\,\rangle}

\newcommand{\LQ}{\Lambda_{QCD}}
\newcommand{\diff}{\partial}

\newcommand{\cT}{{\cal T}}

\newcommand{\mean}[1]{{\langle #1 \rangle}}
\newcommand{\fm}{\mathrm{\,fm}}
\newcommand{\GeV}{\mathrm{\,GeV}}
\newcommand{\MeV}{\mathrm{\,MeV}}
\abstract{
We critically reconsider our recent observation of confining string shrinkage
in pure glue SU(2) lattice gauge theory near the continuum limit. 
Using the advanced numerical techniques we argue that imperfect overlap
with the string ground state and corresponding ambiguities in $T\to\infty$
extrapolation are likely to be the cause of the observed scaling violations.
In particular, even in the limit of large Euclidean times the string shrinkage is
apparent in torelon correlation function, however, in this case the best
relevant overlap we could attain is $\sim 50\%$. To the contrary, when
the ground state is selected properly the string width scales in physical units
being $\approx 0.3\fm$ for $1\fm$ long confining string.
}
\FullConference{8th Conference Quark Confinement and the Hadron Spectrum \\
		 September 1-6 2008\\
		 Mainz, Germany}
\begin{document}
\section{Introduction}
Formation of the confining string between external heavy quark and antiquark
naturally explains the linearly rising heavy quark potential as well as specific stringy
corrections present in it~\cite{str-pot}.  This phenomenon is confirmed in numerous
lattice studies (see, e.g., Refs.~\cite{str}), which however are always limited to
some finite range of UV cutoff (lattice spacing $a$).
However, the internal structure of the confining string
in terms of various gluonic observables is less known, although it is generically
expected and confirmed on the lattice~\cite{str,str-other} that within the flux tube the energy density
is enhanced, while the action density, the magnitude of topological fluctuations
and other quantities are suppressed with respect to the corresponding vacuum expectations.
Note that the local action density difference
\beq\label{delta}
\Delta s = \mean{s}_0 - \mean{s}_{q\bar q} > 0\,, \qquad
s = \tr F^2_{\mu\nu}
\eeq
is  distinguished since its integral matches the heavy quark potential~\cite{ref:sum-rules}.
In the limit of large $q\bar q$ distances $R$ 
the corresponding relation for SU(2) gauge group reads
\beq\label{sumrule}
\int d^2r \, \Delta s ~=~ 2 \sigma \, \beta_f\,, \qquad
\beta_f = - \frac{\diff \ln a }{\diff \beta} \,\,
\stackrel{a\to 0}{\longrightarrow} \,\, \frac{6\pi^2}{22} \,,
\eeq
where the integral is over the flux tube transverse cross section and
$\sigma$ is the string tension. Note that the rigorous sum rule (\ref{sumrule})
says hardly anything about actual action density profile. Nevertheless,
one expects rather generically that  its functional form is to be Gaussian~\cite{str-gauss}
\beq\label{prof}
\Delta s( r, R) ~=~ C(R) \, \exp\{ - r^2 / \delta^2(R)\}\,, \qquad
\delta^2(R) ~=~ \frac{\zeta}{\pi\sigma} \, \ln[R/R_0]\,,
\eeq
where we quoted common to all effective string theories prediction for the squared
width $\delta^2$ of the flux tube, $\zeta$ is model-dependent numerical constant and
$R_0$ is UV cutoff of the effective theory.
True that the string action density profile attracted much attention in
the past~\cite{str}. The summary of these investigations is that
both the Gaussian shape and the string widening, Eq.~(\ref{prof}), are qualitatively confirmed
for large enough quark-antiquark separations $R \gtrsim 0.7\fm$.
Although quantitatively there are points worth to be improved, the above qualitative
picture will hardly change. Thus the question is why we decided to consider the flux
tube profile again and what are the subtle points we hope to shed some light on?

Briefly, the reason reduces to the recent observation that non-perturbative fluctuations,
commonly considered in relation to confinement and known as Abelian monopoles
and center vortices, have a non-trivial ultraviolet
structure~\cite{defects} (see~\cite{viz1,viz2} for comprehensive discussions).
Namely, the excess of non-Abelian action associated with both the monopoles and vortices
is UV divergent, while their number density  perfectly scales towards
the continuum limit.
On the other hand, in Yang-Mills theory all UV divergences must be interpretable within
the field theory itself. Confronting the field theory with the lattice data
one concludes that vacuum defects (monopoles and vortices) are, in a sense,
dual to high orders of perturbation theory (see~\cite{viz1} for details).
For the present discussion it is crucial that for the action density 
the impact of the vacuum defects (or, alternatively, of the ambiguities
of the perturbation theory at high orders) is to be parametrized by an additional
quadratic term in the conventional OPE-based expansion
\beq\label{vdiv}
\mean{s}_0 ~=~ a^{-4} \cdot \left\{\,
\sum_{n=0}^{N} b_n \,\alpha^n_s ~+~ c^{N}_2 \, (a \cdot \LQ)^2
~+~ c_4 \, (a \cdot \LQ)^4 \right\}\,.
\eeq
The necessity of the quadratic correction is quite evident in the context of vacuum defects.
However, from the field-theoretical viewpoint, which is actually implied by (\ref{vdiv}),
a remark have to be added. Namely, Eq.~(\ref{vdiv}) explicitly refers
to a particular N-loop order of perturbative expansion, the divergent reminder of which
is parametrized by the second term. In this respect (\ref{vdiv}) does not follow from
the theory itself, rather it reflects the inability to handle high orders of
$\alpha_s$-expansion and the successful phenomenology of quadratic power corrections
(see Ref.~\cite{q2reviews} and references therein).
In particular, by no means Eq.~(\ref{vdiv}) implies OPE violation.
Consequently, in the field theory settings the quadratic correction cannot be discussed
out of context of the perturbation theory. This is reflected in the explicit $N$-dependence
of the coefficient $c^N_2$: with increasing number of perturbative loops explicitly accounted for
it drops rapidly being (numerically) compatible with zero in the empirical
limit $N\to\infty$~\cite{c2-fit}. Nevertheless, for moderate $N$, when the quadratic term could
be detected, it turns out to be parametrically small~\cite{c2}
\beq\label{estim}
c^N_2 \, \LQ^2 ~\lesssim ~ [50 \MeV]^2\,,
\eeq
which is only an order of magnitude estimation (concrete numbers vary depending
on the details of the measurements).

Given the above uncertain status of the quadratic divergence it is legitimate to ask whether
the issue is at all relevant. The fair answer seems to be affirmative
because we aware of the significance of vacuum defects to QCD vacuum structure.
The discovery of non-trivial UV structure of monopoles/vortices provides a bridge
to the field theory linking them to the question of power divergences. Hence it is worth to consider
Eq.~(\ref{vdiv}) in its generality and to consider the most general  power dependence of
$\mean{s}_0$ upon the UV/IR scales
\beq\label{vpost}
\mean{s}_0 ~=~ c^{(0)}_0 \cdot a^{-4} ~+~ c^{(0)}_2  \,\LQ^2 \cdot a^{-2} ~+~ c^{(0)}_4  \,\LQ^4
\eeq
with some unknown coefficients. True that Eq.~(\ref{vpost}) by itself makes no much sense
because the vacuum action density is not observable anyway. However, it is crucial that
once the quadratic term does not vanish identically no symmetry could ever prevent its dependence
upon the external soft fields. Hence the corresponding VEV $\mean{s}_{q\bar q}$
is to be parametrized analogously with a priori distinct coefficients
$c^{(q\bar q)}_{2,4}$. Then the difference (\ref{delta}) inside the string is given by
\beq\label{adelta}
\Delta s ~=~ c_2  \,\LQ^2 \cdot a^{-2} ~+~ c_4  \,\LQ^4\,,
\eeq
while Eq.~(\ref{estim}) provides an upper bound on $c_2 \LQ^2$.
It is worth mentioning that in terms of vacuum defects Eq.~(\ref{adelta}) with $c_2 \ne 0$
follows directly from the known distribution of monopoles~\cite{str-mono}
 and vortices~\cite{str-vortex} in the vicinity of the flux tube.
Combining (\ref{sumrule}), (\ref{adelta}) one concludes that non-vanishing $c_2$
coefficient manifests itself in peculiar linear shrinkage of the flux tube width
with simultaneous quadratic divergence of the height of the Gaussian profile (\ref{prof})
\beq\label{shrink}
\delta^2 ~\sim~ \sigma \, a^2 / \LQ^2\,, \qquad C(R) ~\sim~ \LQ^2 \cdot a^{-2}
\eeq
in the limit $a\to 0$ (insisting on Gaussian profile seems to be safe at any finite $a$).
Note that this behavior is only asymptotic, an estimation of the relevant lattice spacings
below which (\ref{shrink}) settles in could be obtained from (\ref{estim}) and by noting that
the term $c_4  \,\LQ^4$ is to be of order gluon condensate. Straightforward calculation
shows that the divergence (\ref{shrink}) should be manifest for $a \lesssim 0.03\fm$
and signifies that numerically the status of quadratic divergence is also uncertain.
Indeed, the relevant scale is rather small parametrically and still leaves
slight possibility that the divergence (\ref{adelta}) could have been missed in the previous studies.
The prime purpose of our investigations, which we summarize in this paper, is to study the scaling
properties of the action density distribution within the long confining string.
For technical reasons we restricted ourselves to SU(2) pure gauge theory with Wilson action,
summary of simulation parameters is given in Table~\ref{lattparams}.

\begin{table}
\centerline{\small{
\begin{tabular}{c|c|c|c|c|c}
\multirow{2}{*}{$\beta$} & \multirow{2}{*}{$a, \fm$} & \multirow{2}{*}{$V_{lat}$, $V_{phys}$} &
\multicolumn{2}{c|}{Multilevel: $\Delta$ and $N_{upd}$} & Multichannel, \\
 & & & Torelons, $N_{lev}=2$ & Wilson loops, $N_{lev}=3$ & $N_{smearing}$ \\ \hline
2.5046 & 0.081 & $28^4$, $[2.3\fm]^4$  & $(14)$, $(4000)$ & -- & -- \\ \hline
2.5515 & 0.072 & $28^4$, $[2.0\fm]^4$  & $(14)$, $(3500)$ & -- & -- \\
       &       & $32^4$, $[2.3\fm]^4$  & $(16)$, $(5000)$ & $(4, 8)$, $(12, 300)$   & $(50,65,80)$   \\ \hline
2.6800 & 0.047 & $36^4$, $[1.7\fm]^4$  & $(18)$, $(4500)$ & -- & -- \\
       &       & $40^4$, $[1.9\fm]^4$  & $(20)$, $(5000)$ & $(5, 10)$, $(12, 600)$  & $(150,180,210)$ \\ \hline
2.7600 & 0.034 & $48^4$, $[1.8\fm]^4$  & $(24)$, $(5000)$ & $(6, 12)$, $(12, 1200)$ & $(180,210,240)$ \\ \hline
\end{tabular}}}
\caption{Ensembles used in our study. Multilevel notations (fourth column) closely follow~\cite{meyer}.
APE smearing (fifth column) had been performed  with $\varepsilon$ parameter $0.25$. }
\label{lattparams}
\end{table}

Our first measurements \cite{self} confirmed the presence of the quadratic
term, Eq.~(\ref{adelta}), with magnitude being in agreement with (\ref{estim}).
However, we had never been quite satisfied with this result for the following reasons:\\
1. Ground state separation (ground state overlap). \\
String creation/annihilation operators used in \cite{self} were not optimal, the relevant overlaps
with the ground state of heavy $q\bar q$-pair were of order $80\div 90\%$. This required us to make rather
complicated fit to $T \to \infty$ thus introducing systematic biases the magnitude
of which we were unable to fairly estimate. This issue is addressed in Section~\ref{WlPl}. \\
2. Limit of large $q\bar q$ separations.\\
Ultimately, we're interested in the physics of long confining string, $R \gg R_0 \approx 0.7 \fm$,
when the $q\bar q$ separation $R$ greatly exceeds the string formation length $R_0$.
With Wilson loops (or Polyakov lines) it is hardly possible to go far beyond $R_0$ due
to the technical limitations, hence the string profile is disturbed by singular point-like sources
at its end-points. Again some phenomenology had been applied in \cite{self}
leading to the same question of systematic errors.  Ideally, the point-like sources
have to be removed, which leads us to consider the torelon correlation function (Section \ref{torelons}).

An overall conclusion of the present study is that it is the overlap problem which is the most
crucial to get the physics right. In particular, when the proper selection of string
ground state is ensured and its $T=\infty$ asymptotic structure is directly seen
the effect of string shrinkage disappears.
To our surprise the string profile turned to be extremely sensitive in this respect, contrary
to, e.g., heavy quark potential in which we never observed scaling violations.
Therefore in string profile studies it is absolutely necessary to get rid of excited states
contamination prior to making any conclusions on the flux tube internal structure.

\section{Torelons}
\label{torelons}

Technically, the torelon operator is similar to Polyakov line: it is the gauge holonomy
around the spatial period of toroidal lattice in a particular time slice
(see, e.g., Ref.~\cite{tor} and references therein).
Physically, it is quite different however: torelon creates confining string closed through
the periodic boundary conditions and stable for topological reasons.
The string exists without point-like sources, hence the torelon correlation function is best
suited to study the flux tube internals. Normalized torelon correlator allows
usual spectral decomposition $\cT(t) ~=~ \sum \, c_i \, e^{-m_i t}$, $\cT(0) = 1 $,
where masses $m_i$ depend upon the spatial lattice extent $L$ and have
the same leading asymptotic $m_i \approx \sigma L$.

\begin{figure}
\centerline{\psfig{file=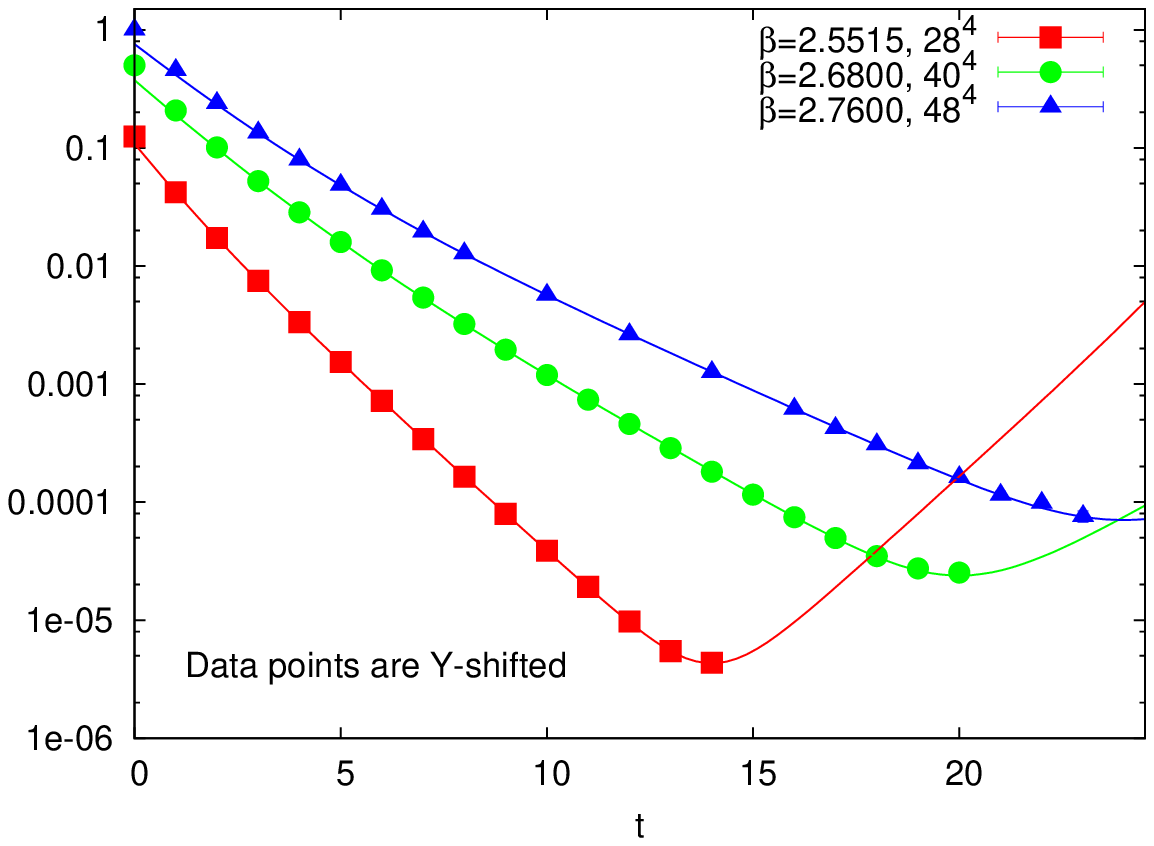,width=0.5\textwidth,silent=}
\psfig{file=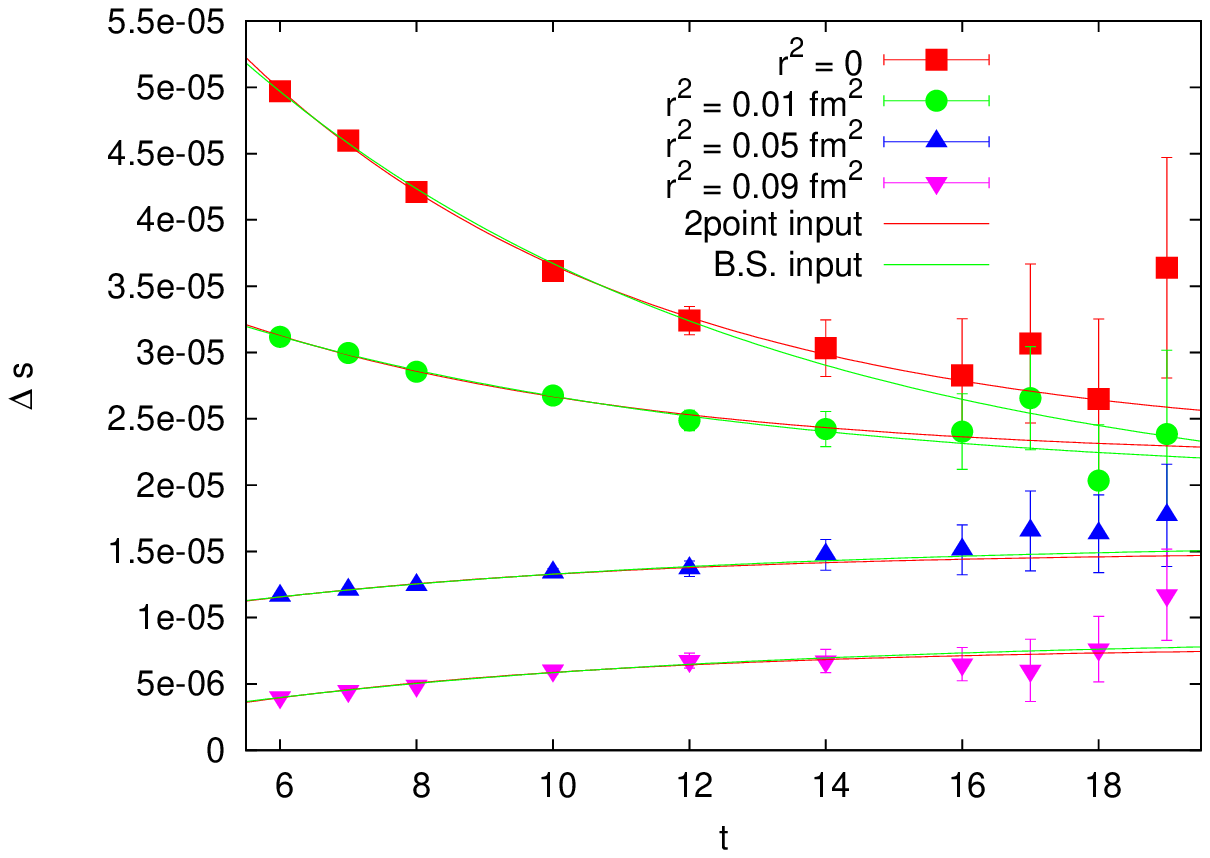,width=0.5\textwidth,silent=}}
\caption{Left: torelon correlator at various spacings/volumes and two-exponential
fits  $\cT(t) ~=~ \sum \, c_i \, e^{-m_i t}$, $c_0\approx 0.4$, $c_1\approx 0.55$.
Right: examples of $t\to\infty$ fits with gap fixed
either from $\cT(t)$ or from bosonic string theory (B.S.), $\beta=2.760$, $48^4$.}
\label{fig:tor-mass}
\end{figure}

In order to study the string profile with torelons it is necessary
to have precise data for the torelon correlation function at large Euclidean times.
This is due to the local action density insertion: for instance, we used clover-like
sum of all neighboring to given point plaquettes, hence the minimal Euclidean time we
could consider is $t=4$ (in fact, it has to be larger for asymptotic exponential falloff
to take place). Thus the use of modern multilevel technique~\cite{mlevel,mlevel2} is mandatory.
In usual multilevel approaches it is the time axis which is divided into slabs, so that
at large Euclidean times torelon creation/annihilation operators belong to different
time slabs. Our experience showed, however, that in order to obtain desired precision
in string profile, the 2-point correlator should be known with by order of magnitude
better accuracy than it is available with usual scheme. Therefore, we utilized another
approach in which the torelon direction is divided among different slabs. The ground
state enhancement is done as usual in each slab, however, it is evident that our scheme
does not allow to get optimal overlap with the string ground state: the more slabs are introduced
the worse is overlap.
Indeed, in each slab one considers straight smeared spatial holonomies with end-points fixed
at the boundaries.  Upon averaging the string wavefunction is constructed as superposition
of various paths, which are required, however, to pierce prescribed boundary points.
Effectively at each boundary one introduces a node of the wavefunction thus reducing the ground
state overlap.
Minimal number of time slabs with the 2-level scheme is two and the corresponding overlap
is about 50\%.
Despite of its smallness we hoped to eliminate exited states contamination by attaining
the limit of large Euclidean times, which is feasible with multilevel strategy.

Sample of data points obtained for the torelon correlator is shown on Fig.~\ref{fig:tor-mass} (left).
Analyzes reveal that we see a superposition of two states
with approximately equal amplitudes. Mass extraction is done via two-exponential fit,
lower $t$-limit of which was varied until lighter mass stabilizes.
Masses extracted this way are in qualitative agreement with the literature~\cite{tor},
however, quantitatively they appear to be systematically slightly larger.
This is because we are obliged to consider 2-point correlator, not the wall-wall correlation function, 
and hence inevitably get an admixture of non-zero lowest spatial momentum states.
Thus the masses extracted from 2-point function could only be used as an approximation
to the true spectral gap in the torelon channel. 

Unfortunately, our hope to observe exclusively the ground state profile at largest available times
failed for small transverse distances (Fig.~\ref{fig:tor-mass}, right).
Indeed, for $r^2 \lesssim 0.03 \fm^2$ $t$-dependence of $\Delta s(t, r^2)$ is apparent,
while at larger $r^2$ an approximate $t$-constancy of the profile seems to settle
at $t \gtrsim 0.5 \fm$. Therefore, it is necessary to make $t\to\infty$ extrapolation
for which the leading finite-$t$ correction looks like single exponent
with gap being fixed in term of the spectral gap in the torelon channel.
However, besides the imprecise knowledge of the gap there is an additional limitation:
due to imperfect overlap with the ground state single exponent might not be adequate
especially at small $r$.
In view of these shortcomings various fits were tried out:
1. Single exponent fits with gap fixed either from the 2-point function or
from the bosonic string theory prediction (they are illustrated  Fig.~\ref{fig:tor-mass}, right).
2. Fit by constant in $t$-range where data points come to plateau, it was
used for qualitative purposes only.
3. Cumulative fit to all available $t$, $r^2$ points
with single exponential ansatz and treating the gap as free parameter.
To ensure the validity of single exponent approximation the points
too close to the string axis were excluded, the corresponding cut on $r$
is obtained from previous fits. 
Additionally, we made seemingly well grounded assumption
that the string profile at $t=\infty$ is to be Gaussian.
This greatly reduces the dimensionality of the problem and allows much better precision.

\begin{figure}
\centerline{\psfig{file=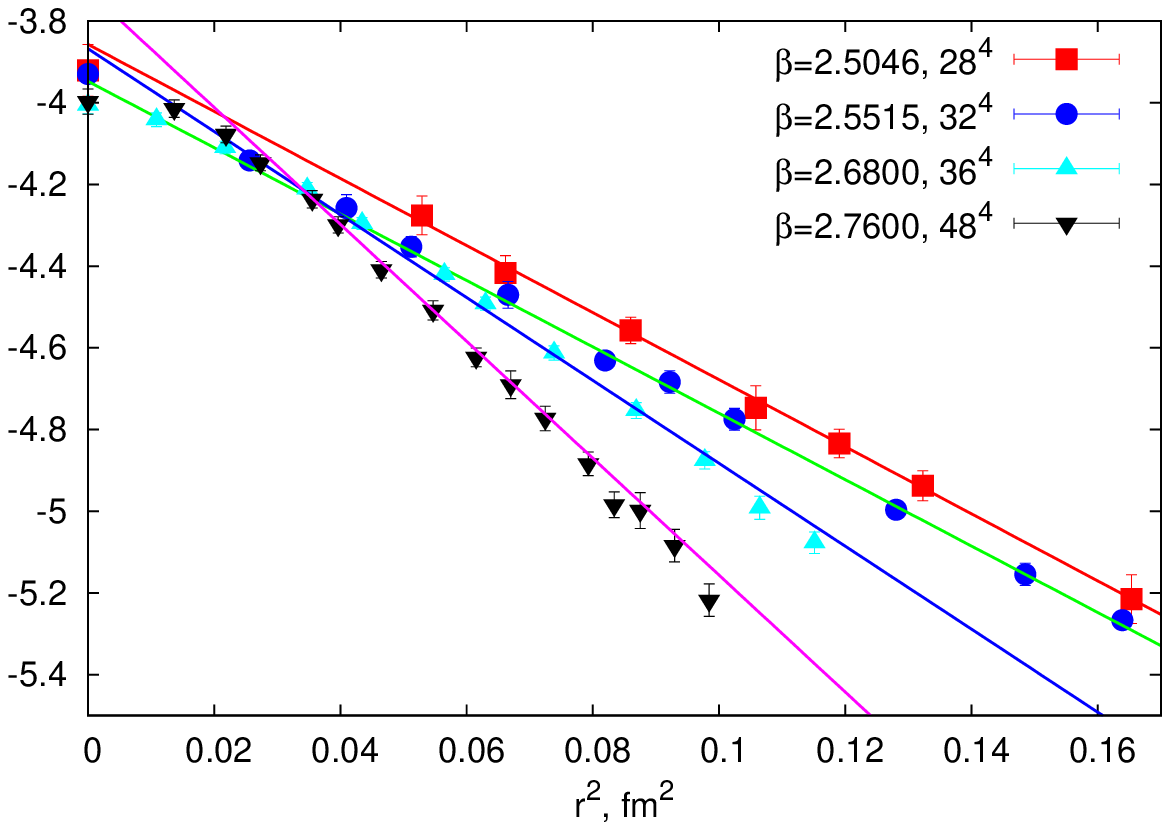,width=0.5\textwidth,silent=}
\psfig{file=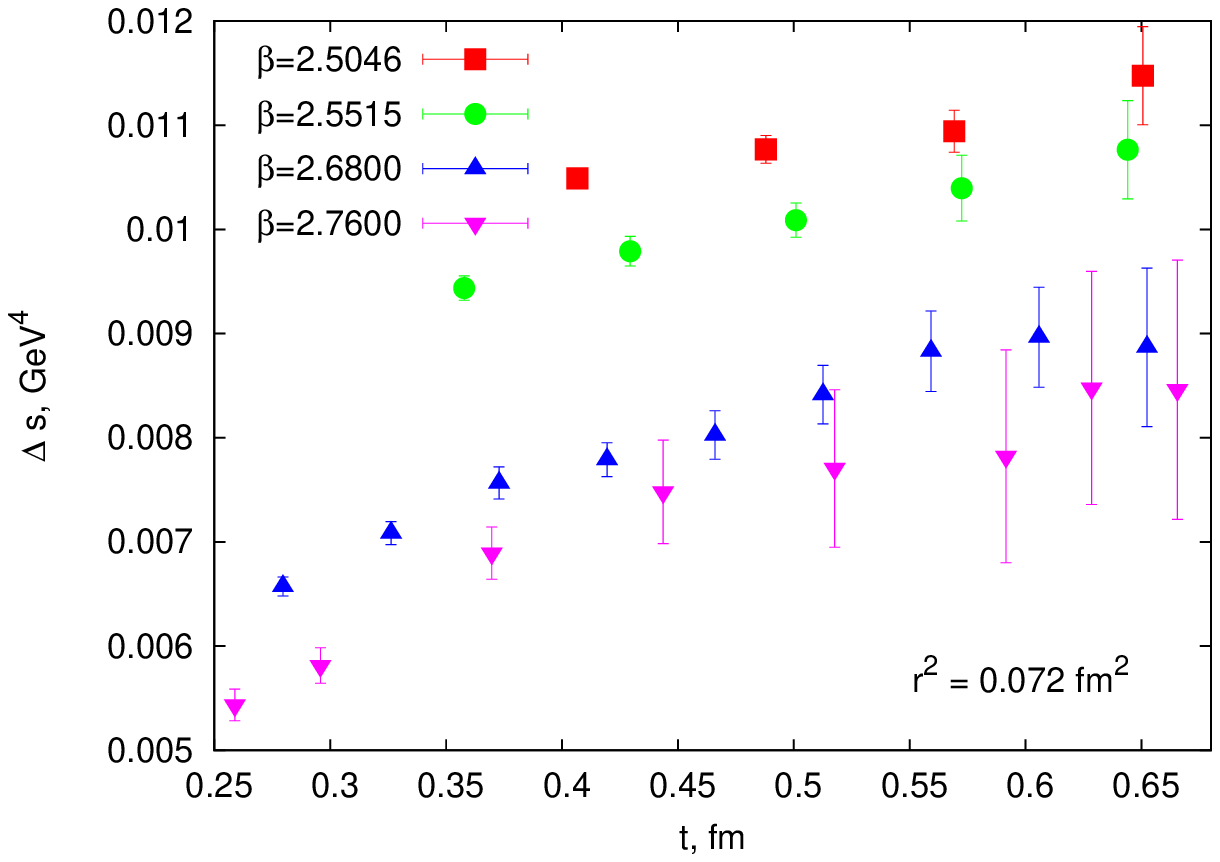,width=0.5\textwidth,silent=}
}
\caption{Left: $\ln\left(\Delta s(t=\infty)/[1\GeV]^4\right)$ vs. $r^2$ at various lattice
resolutions obtained with fixed gap fits (points) and cumulative one (solid lines).
Right: $\Delta s(t)$ at $r \approx 0.27 \fm$ at different lattice spacings.}
\label{fig:tor-fit}
\end{figure}

Summary of the results obtained with various fitting procedures is presented on
Fig.~\ref{fig:tor-fit}, left panel. Briefly, all the above fits appear to be adequate and consistent 
with each other except
for the region $r^2 \lesssim 0.03 \fm^2$ at smallest lattice spacing, where fixed gap fits
notably deviate from Gaussian.
Treating the discrepancies at small $r$ as a consequence of the imperfect overlap
and the enforcement of single exponential anzatz, we conclude 
that regardless of actual $t\to\infty$ extrapolation the string width seems to drop
with diminishing lattice spacing.
In fact, the apparent string shrinkage could be identified directly by looking on the data
at roughly the same transverse distance $r \approx 0.27 \fm$
and various lattice resolutions (Fig.~\ref{fig:tor-fit}, right).
Furthermore, the magnitude of the quadratic divergence turns out to be
$c_2 \LQ^2 \approx [30\,\MeV]^2$ and essentially the same number follows from
the consideration of the string profile height.
Although the above treatment seems to be self-consistent,
it is crucial to investigate the systematic biases of the ground state
extraction, to which we turn next.

\section{Wilson loops}
\label{WlPl}

\begin{figure}
\centerline{\psfig{file=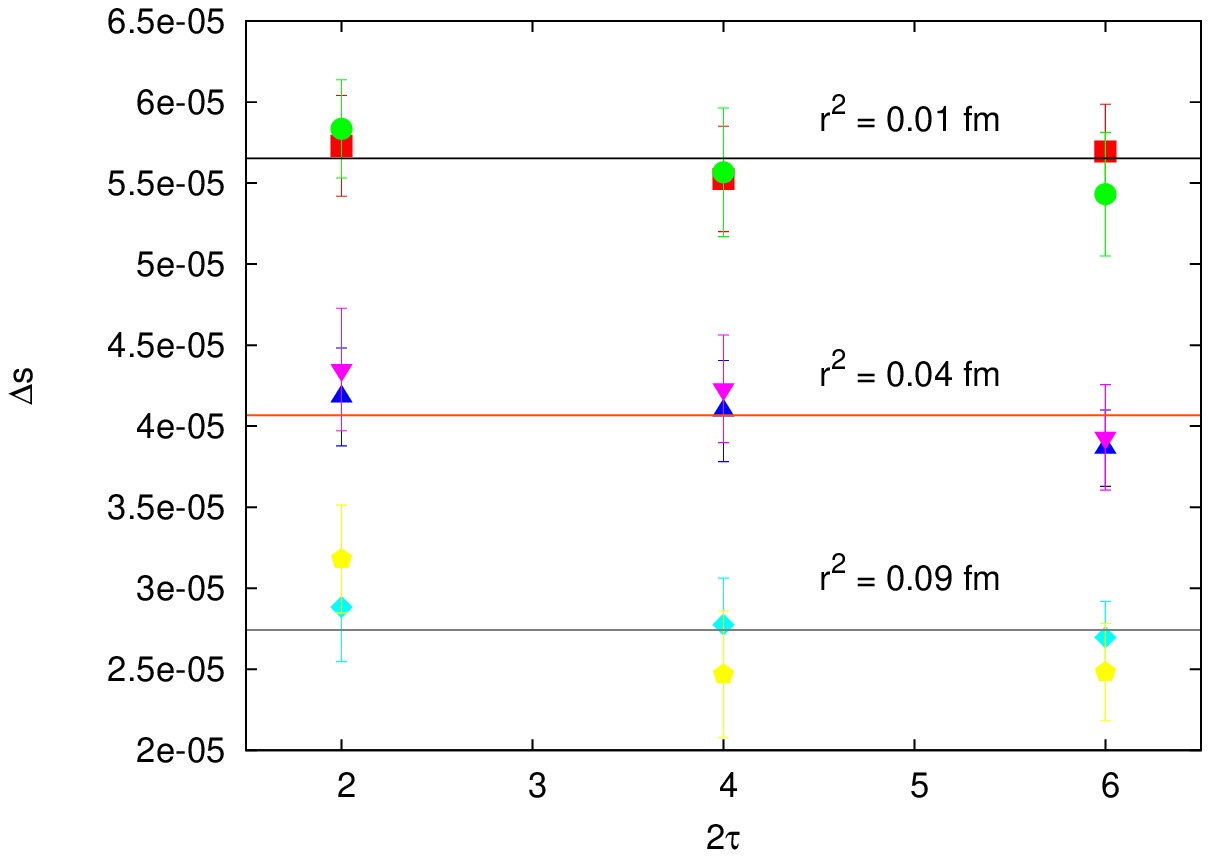,width=0.5\textwidth,silent=}
\psfig{file=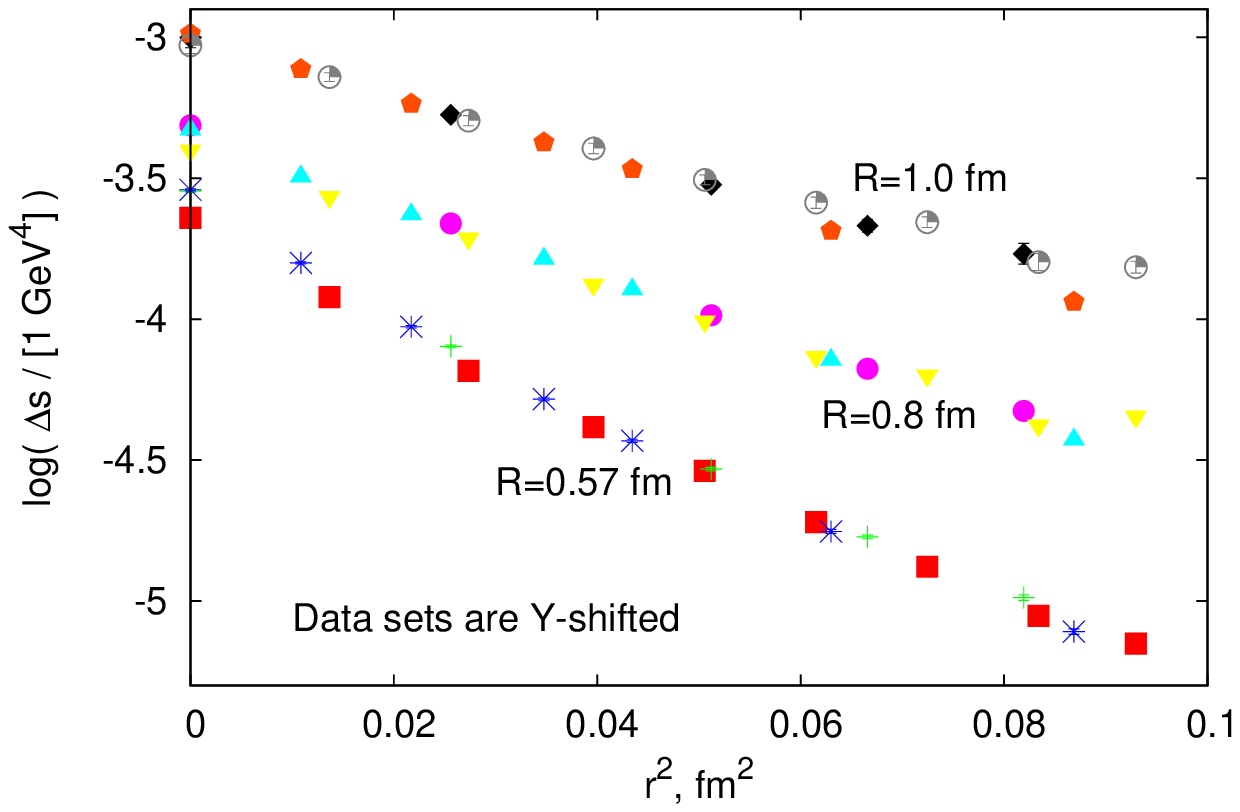,width=0.5\textwidth,silent=}}
\caption{Left: $\Delta s$  vs. $\tau$ at $(t_0, t) = \{(8,18), (8,22)\}$,
$R = 0.93\fm$ and different $r^2$ ($\beta=2.680$, $40^4$) with the corresponding
fits to constant. Right: flux tube profiles at various lattice resolutions. }
\label{fig:wl1}
\end{figure}

It seems that the only reliable way to estimate the precision of $t\to\infty$ extrapolation
is to get rid of it and to ensure that only the ground state propagates in the considered channel.
To this end we combined the multilevel (actually, 3-level) scheme of Wilson loops calculation
with the special ground state selection technique conventionally known as
multichannel~\cite{mchanell}. In essence it reduces to choosing the set $\{\ket{i}\}$,
$i = 1,...,N_{st}$ of string states and solving the generalized eigenvalue problem
(GEVP)
\beq\label{gevp}
W_{ij}(t,R) \, e^{(\lambda)}_j ~=~ \lambda \, W_{ij}(t_0,R) \, e^{(\lambda)}_j\,.
\eeq
Here $W_{ij}(t,R)$ is $t \times R$ Wilson loop correlator evaluated for
the states $\ket{i}$, $\ket{j}$, which we prepared using different number
of APE smearing sweeps with the same smearing parameter (see Table\ref{lattparams}).
The best available approximation to the ground state corresponds to the
eigenvector $e^{(0)}$ with maximal eigenvalue $\lambda^{(0)}$. For the
string profile studies this approach applies literally and leads to
\beq\label{delta-gevp}
\Delta s = \mean{s}_0 ~-~ e^{(0)}_i [Ws]_{ij} e^{(0)}_j/\lambda^{(0)}\,,
\eeq
where $[Ws]_{ij}$ correlation matrix is similar to $W_{ij}$,
but with local action density operator inserted at time-like [transverse]
distances $\tau$ [$r$] from the loop geometrical center (longitudinal shift
was kept vanishing). Technically, our approach is similar to that of Ref.~\cite{mlevel2}:
at the lowest (deconfining) level of our 3-level scheme only the products of time-like open holonomies
(possibly with action density insertions) were averaged. At the next (confining)
level these were accumulated and open staple-like segments of Wilson loops were
calculated as well (to reduce the statistical errors the space-like pieces of the loops were
required to be at least $l_{cr}/2$ distance away from the slab boundaries,
$l_{cr}$ being the inverse temperature of deconfinement transition at fixed boundaries).
Finally, at the upper-most level the averaged open holonomies
and their tensor products were combined to obtain the correlation matrices $W_{ij}$, $[Ws]_{ij}$.
Note that with definition (\ref{delta-gevp}) $\Delta s$ depends upon five parameters,
$(t_0, t, \tau, R, r^2)$, where the triple $(t_0, t, \tau)$  is used exclusively to ensure
the proper ground state selection.

The multichannel approach with three trial states allows to significantly
increase the ground state overlap. In all cases it was at least 0.99
differing from unity in the third digit well within the statistical errors.
As a consequence the dependence $\Delta s(t_0, t, \tau)$ tuned to be trivial as is
exemplified for $\beta=2.680$, $40^4$ data on Fig.~\ref{fig:wl1} (left), where
$\Delta s$ at $R=0.93\fm$ is plotted as a function of $\tau$ with two choices of
GEVP window $(t_0, t) = \{ (8, 18), (8, 22) \}$ and various transverse distances.
The straightforward constant fit
gives then the $t=\infty$ asymptotic of the string profile, which is presented on
Fig.~\ref{fig:wl1}, right. To our surprise, the points obtained at various lattice
resolutions fall, in fact, on the top of each other with no sign of scaling violations.
The only conclusion we could draw is that the undertaken $t\to\infty$
extrapolations are in fact ambiguous despite of their apparent self-consistency.
Therefore the previous claim of the string shrinkage is void and at least within
the precision of our data the quadratic divergence is not seen in the action density
difference (\ref{delta}). 

\begin{figure}
\centerline{\psfig{file=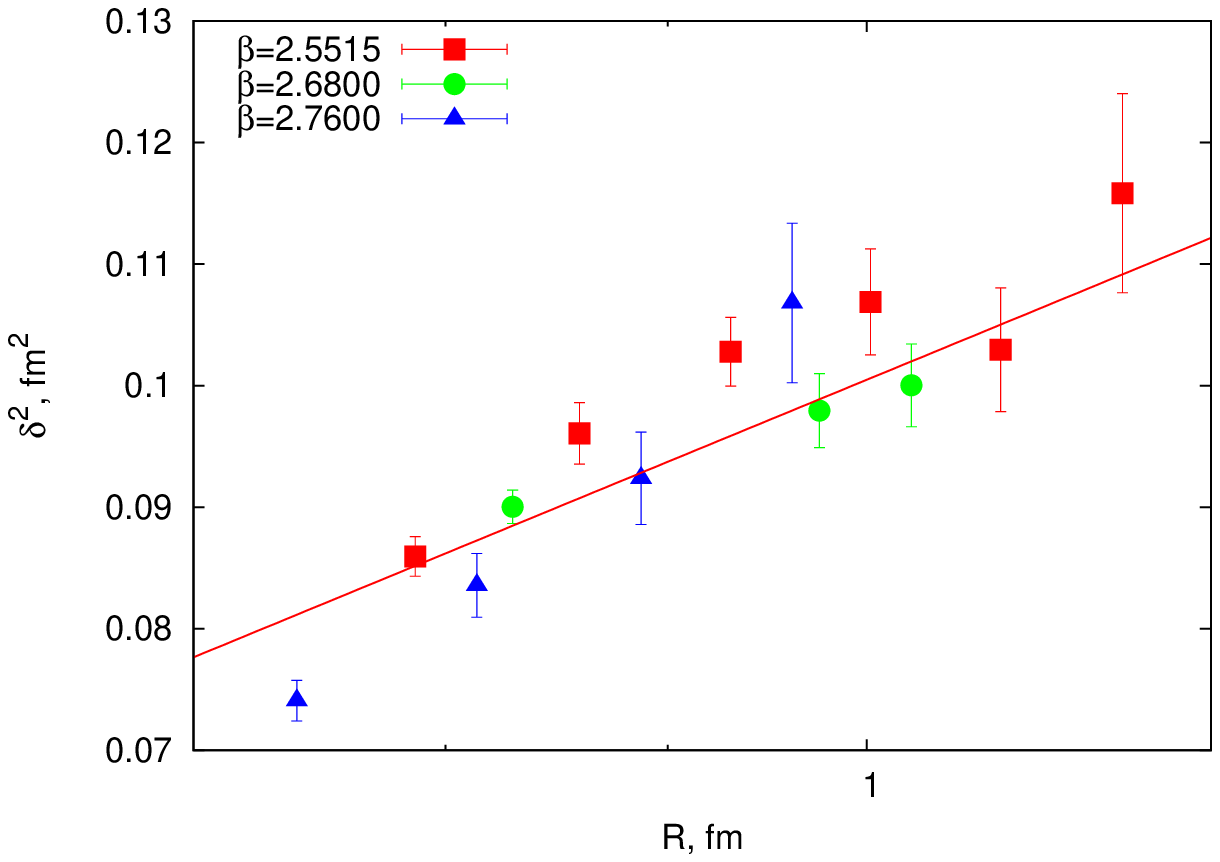,width=0.5\textwidth,silent=}
\psfig{file=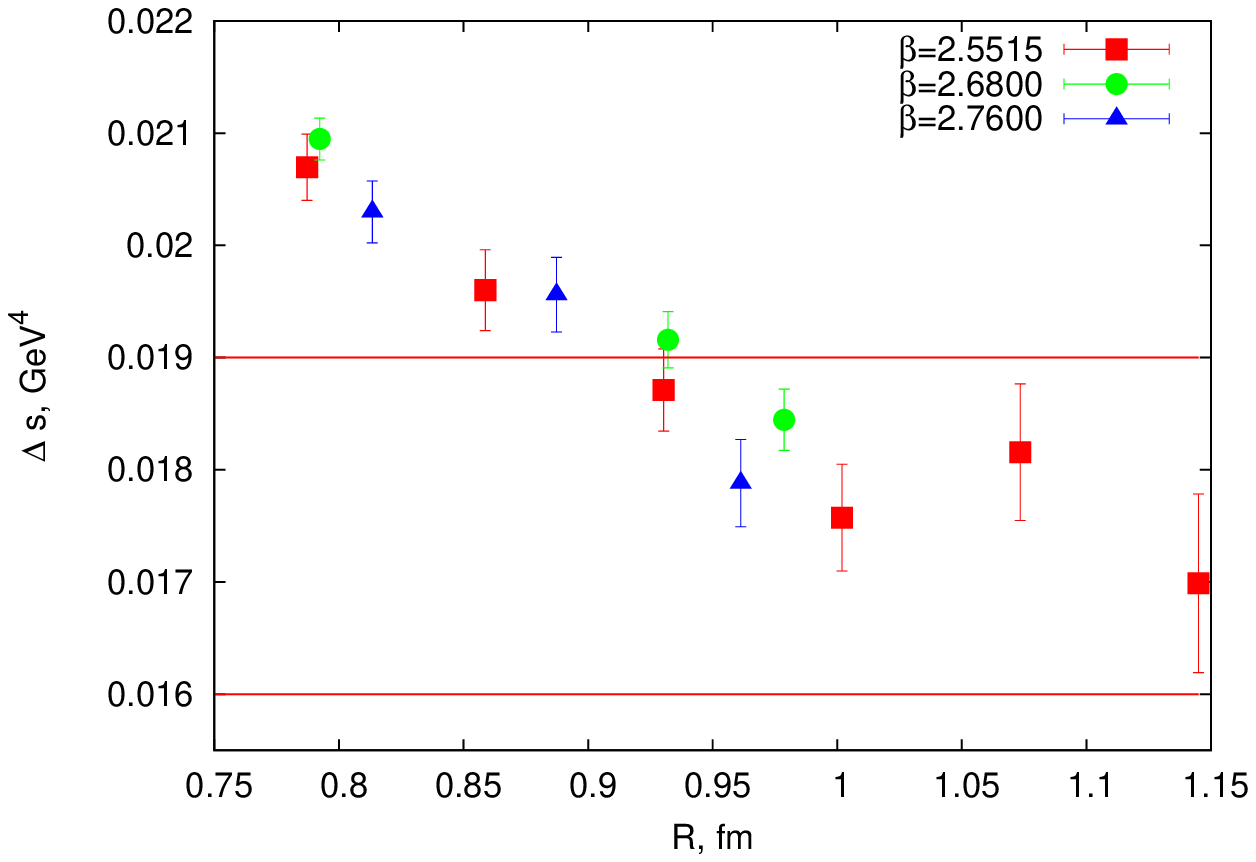,width=0.5\textwidth,silent=}}
\caption{Left: Squared string width $\delta^2$ versus distance $R$ plotted in logarithmic scale.
Straight line illustrates the bosonic string theory prediction (1.3), $\zeta = 1$, with fitted
parameter $R_0 \approx 0.2\fm$. Right: height of the Gaussian profile, $\Delta s(r^2 = 0)$,
as a function of $R$, solid lines represent the rough upper/lower estimates
of the gluon condensate. }
\label{fig:wl2}
\end{figure}

Once the claim of linear vanishing of the string width with the lattice spacing had been
abandoned it is worth to consider the dependence of $\delta^2$ upon the distance $R$
(Fig.~\ref{fig:wl2}, left). According to (\ref{prof}) it is expected to be logarithmically increasing
function and indeed Fig.~\ref{fig:wl2} qualitatively confirms this. 
It is also apparent that the data does not allow quantitative conclusions, only order of magnitude
estimates could be given. To this end we tried the logarithmic fit of the type (\ref{prof}).
Although the data prefer $\zeta$ value slightly above one,
the difference is insignificant. Thus we fixed $\zeta$ to unity and evaluated the $R_0$ parameter,
the optimal value of which turned to be $R_0 \approx 0.2\fm$. 

Finally, let us remark that for $\approx 1\fm$ long confining string the height of its action
density profile is about $0.018 \GeV^4$ (Fig.~\ref{fig:wl2}, right). Numerically, this is
very similar to the magnitude of the gluon condensate in SU(2) gluodynamics~\cite{gl}.
We qualitatively conclude therefore that gluon condensate vanishes on the symmetry axis
of well developed confining string.

\section{Conclusions}
In this paper we critically reviewed the observation of the confining string
shrinkage in the continuum limit of pure glue SU(2) lattice Yang-Mills theory.
The aim was to address the problem of systematic biases coming from the imperfect
ground state determination and from the string profile distortions due to
the presence of singular point-like sources at its end-points. The second
question is proved to be inessential: in the case of torelon correlator, where the flux tube
exists without point-like sources, the string shrinkage had been also identified.
However, addressing the issue of excited states contamination
we discovered that it is a very delicate and perhaps not a well defined problem
of taking the limit of infinite Euclidean times. Despite of apparent self-consistency
of considered $t\to\infty$ extrapolations they lead to incorrect results.
Namely, in the case of Wilson loops it is possible to construct almost perfect
approximation to the string ground state, the shape of which is in fact independent
upon the Euclidean time extent. Then the effect of string shrinkage completely
disappears proving that it is an artifact of improper ground state selection.
As a byproduct of our analyzes we qualitatively confirm the validity of the effective
bosonic string theory description of the string profile and argue that the gluon
condensate vanishes within the well developed confining flux tube.

\end{document}